\documentclass{elsart}

\usepackage{graphicx,natbib,amssymb,lineno}
\usepackage{epstopdf}
\usepackage{bm}

\makeatletter
\def\elsartstyle{%
    \def\normalsize{\@setfontsize\normalsize\@xiipt{14.5}}
    \def\small{\@setfontsize\small\@xipt{13.6}}
    \let\footnotesize=\small
    \def\large{\@setfontsize\large\@xivpt{18}}
    \def\Large{\@setfontsize\Large\@xviipt{22}}
    \skip\@mpfootins = 18\p@ \@plus 2\p@
    \normalsize
}
\@ifundefined{square}{}{}
\makeatother

\usepackage{epsfig}

\begin{document}

\begin{frontmatter}

\title{Driven Ratchets for Cold Atoms}

\author{Ferruccio Renzoni}

\address{Departement of Physics and Astronomy, University College London,
Gower Street, London WC1E 6BT, United Kingdom}

\begin{abstract}
% Text of abstract
Brownian motors, or ratchets, are devices which ``rectify" Brownian
motion, i.e. they can generate a current of particles out of unbiased
fluctuations. The ratchet effect is a very general phenomenon which applies
to a wide range of physical systems, and indeed ratchets have been realized
with a variety of solid state devices, with optical trap setups as well as
with synthetic molecules and granular gases. The present article reviews
recent experimental realizations of ac driven 
ratchets with cold atoms in driven optical lattices. This is quite an 
unusual system for a Brownian motor as there is no a real thermal bath, 
and both the periodic potential for the atoms and the fluctuations are
determined by laser fields. Such a system allowed us to realize experimentally
rocking and gating ratchets, and to precisely investigate the
relationship between symmetry and transport in these ratchets, both
for the case of periodic and quasiperiodic driving.

\end{abstract}

\begin{keyword}
Ratchets \sep cold atoms \sep optical lattices
% keywords here, in the form: keyword \sep keyword
\PACS 05.45.-a  \sep 42.65.Es  \sep 32.80.Pj
% PACS codes here, in the form: \PACS code \sep code
\end{keyword}

\end{frontmatter}

\tableofcontents

%*******************************************************************
%* 1. INTRODUCTION
%*******************************************************************

\section{Introduction}
\label{}

Brownian motors, or ratchets, are devices which rectify fluctuations, 
turning in this way unbiased Brownian motion into directed diffusion,
in the absence of net applied bias forces.

The concept of ratchet was initially introduced to point out the strict 
limitations on directed transport at equilibrium imposed by the second
principle of thermodynamics \citep{feynman}. Ratchets have then been 
attracting growing attention in different communities for the number of 
applications: from particle separation, to the modelling of molecular 
motors, and to the realization of novel types of electron pumps, just to
name a few. Recent reviews (\cite{reimann02}, \cite{rmp09})
provide a detailed account of the theoretical
work relevant to the ratchet effect, the experimental realizations in 
many different fields and related practical applications.

The present article reviews recent realizations of driven ratchets for cold 
atoms. A previous review \citep{renzoni} summarized the experimental
work at that time. In these cold atom systems, light fields create both a
periodic potential for the atoms, and introduce fluctuations in the atomic
dynamics. Appropriate
ac drivings can also be introduced. The so realized driven ratchets allowed
us to experimentally demonstrate many of the characteristic features of
ratchets, as for example current reversals. The precise control on the 
ac drivings also allowed us to investigate from an experimental point of 
view the relationship between symmetry and transport, which is the 
essential element for the understanding of the operation of a ratchet.

This review article is organized as follows. In Section \ref{sec:generalities}
the concept of ratchet is introduced, and two early proposals of ratchets 
discussed: the flashing and the rocking ratchets. In Section 
\ref{sec:symmetries} the important role that symmetries play in the operation 
of a ratchet device is discussed. The symmetry analysis, initially introduced 
from a general
point of view, will then be specialized to a periodically and quasiperiodically
driven rocking ratchet, and to a gating ratchet. In Section \ref{sec:coldatoms} recent
experimental realizations of driven ratchets for cold atoms are reviewed. 
After introducing the main features of dissipative optical lattices, 
specific experimental realizations of driven ratchets for cold atoms are
examined: a periodically and a quasiperiodically rocking ratchet, and the 
gating ratchet.  Finally, in Sec. \ref{sec:outlook} possible future directions
of research in cold atom ratchets are discussed.

\section{Ratchets: generalities} \label{sec:generalities}

Brownian motors are devices which produce a current out of unbiased
fluctuations. Strict limitations on the operation of a ratchet are 
imposed by the second principle of thermodynamics, which rules out 
the possibility of producing a current at thermodynamic equilibrium.
Thus, the effective generation of a current requires the system to be 
driven out of equilibrium. We will now examine how this is implemented 
in two specific cases of ratchet devices: the flashing and the rocking 
ratchets.

\subsection{The flashing ratchet}

\begin{figure}[ht]
\begin{center}
\mbox{\epsfxsize 4.5in \epsfbox{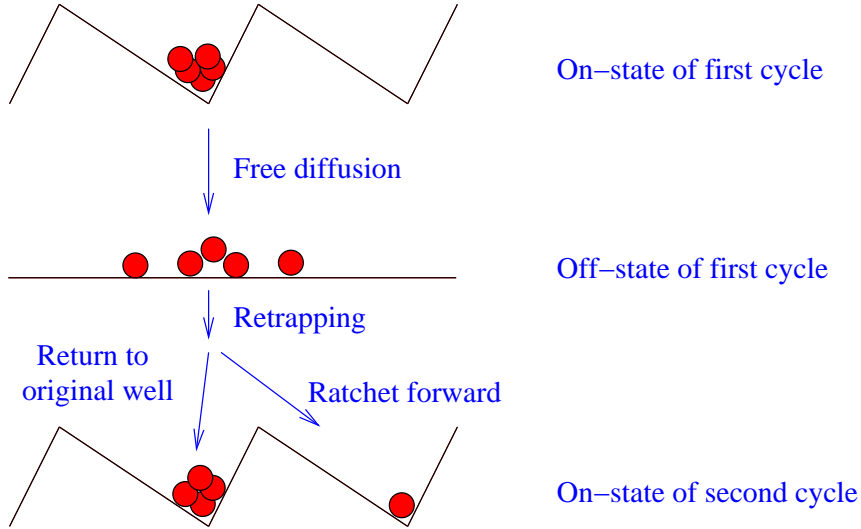}}
\end{center}
\caption{Working principle of the flashing ratchet.}
\label{fig1}
\end{figure}

Consider a sample of Brownian particles in a (static) asymmetric periodic 
potential. The second  principle of thermodynamics rules out the 
possibility of directed motion. However, things are very different if the 
potential is "flashed", i.e. if it is turned on and off repeatedly, either
periodically or randomly \citep{comptes,juliette}. This is sufficient to set 
the Brownian particles into directed motion, due to the mechanism illustrated
in Fig.~\ref{fig1}.

%%%%%%%%%%%%%%%%%%%%%%%%%%%%%%%%%%%%%%%%%%%%%%%%%%%%%%%%%%%%%%%%%%%%%%%%%%%
%%%%%%%%%%%%%%%%%%%%%%%%%%%%%%%%%%%%%%%%%%%%%%%%%%%%%%%%%%%%%%%%%%%%%%%%%%%

Consider an initial situation with the potential turned on and the Brownian
particles localized at the bottom of a given well. Then the potential is
turned off, and the Brownian particles will symmetrically diffuse in space.
Then the potential is turned on again, and the Brownian particles are retrapped
in both the original well and in a few neighbouring ones. However, as the
potential is asymmetric the retrapping will lead to an asymmetric situation,
with the number of particles trapped in the wells at the left of the original
well different from the number of particles trapped in the wells at the right
of the starting location. Indeed it is clear from Fig.~\ref{fig1} that the
wells closer to the ``steep wall" of the starting well will collect more particles
during the retrapping phase. In this way the center of mass of the particle
cloud will move, and directed motion is thus obtained.

It is important to point out why the operation of the flashing ratchet does
not violate the second law of thermodynamics. This is because work is done
on the system while turning on the potential. Thus, although fluctuations
are rectified and a current is generated, this does not imply that work has
been extracted out of just one heat source as some additional work was
necessary to turn on the potential. Therefore the second law of
thermodynamics is not violated.

\subsection{The rocking ratchet}

In the {\it rocking} ratchet \citep{magnasco,adjari94,bartussek,doering}, 
particles in a periodic
asymmetric potential experience also an applied ac force. The applied 
force, which is zero-average and time-symmetric, drives the system out
of equilibrium. As a results of the symmetry-breaking anisotropy of the 
potential, a net current of particles can thus be generated. The same effect 
can be obtained for a spatially symmetric potential and a temporally
asymmetric drive \citep{mahato,luczka,chialvo}. A bi-harmonic force is 
a popular choice for a time asymmetric drive,
with the time-symmetry of the drive controlled by the relative phase
between harmonics \citep{flach00,flach01,super,acta}. In the latter case of 
symmetric potential, and multi-harmonic driving, the underlying rectification 
mechanism can be traced back to harmonic mixing \citep{marchesoni}.

Rocking ratchets, and more in general ac driven ratchets, are the central topic 
of the present review. Therefore, in the following the relationship between 
symmetry and transport will be examined in detail for these ratchets.  

\section{Symmetry and transport in ac driven ratchets} \label{sec:symmetries}

The operation of a ratchet requires an out-of-equilibrium set-up, and the
breaking of the symmetries which would otherwise prevent directed motion.
This Section reviews the symmetry analysis for the specific case of ac driven 
ratchet, as derived by \cite{flach00},\cite{flach01},\cite{super},\cite{acta}.

\subsection{General considerations}\label{sec:general}

We consider a Brownian particle in a spatially periodic potential $U$ of period 
$\lambda$. A time-dependent driving force $F$, of zero mean, is applied to the 
particle.  The Langevin equation for the
particle of mass $M$ is:
\begin{equation}
M\ddot{x}+\gamma\dot{x}=-U'(x) + F(t)+\xi(t)~,
\label{eq:langevin}
\end{equation}
where $U'(x)$ denotes the first derivative of the function $U$. 
Here $x$ is the position of the particle at the time $t$, and $\gamma$ and
$\xi$ are the damping coefficient and a stationary Gaussian noise respectively.

Following a standard procedure in the symmetry
analysis of ratchet devices \citep{flach00,flach01,super,acta}, we aim to
determine the conditions for the Langevin equation, Eq.~(\ref{eq:langevin}),
to be invariant under the following symmetries
\begin{eqnarray}
\hat{S}_1:~ x\to -x+x',~~~t\to t+\tau \label{eq:s1}\\
\hat{S}_2:~ x\to x+\chi,~~~t\to -t+t' \label{eq:s2}
\end{eqnarray}
with $x',t'$, $\tau$ and $\chi$ constants.  These are the tranformations which
map a trajectory $\{x(t,x_0,p_0),p(t,x_0,p_0)\}$, with $x_0,p_0$ the initial
position and momentum, into one with opposite momentum. The invariance of
the Langevin equation under $\hat{S}_1$ and/or $\hat{S}_2$ then prevents
directed motion.

\subsection{The periodically driven rocking ratchet}\label{sec:periodic}

Whether $\hat{S}_1$, $\hat{S}_2$ are symmetries of the system depends on the
specific form of $U(x)$ and $F(t)$.  Throughout the present review, we consider 
only the case of a spatially symmetric periodic potential $U(x+\chi)=U(-x+\chi)$, 
where $\chi$ is a constant. This is the case relevant to the experimental
realizations reviewed in this work, with the symmetry of the system controlled
by the ac driving. In this Section, we examine the case of a periodic driving 
$F(t)$, of period $T$.  Following the notations of \cite{flach00}, we say that
$F(t)$ possesses $\hat{F}_s$ symmetry if $F(t)$ is invariant under time
reversal, after some appropriate shift:
\begin{equation}
F(t+\tau)=F(-t+\tau)~.
\end{equation}
Moreover, if $F(t)$ satisfies:
\begin{equation}
F(t)=-F(t+T/2)
\end{equation}
we say that $F$ possesses the $\hat{F}_{sh}$ shift-symmetry.

We first consider the dissipationless case, which will then be extended to 
include weak dissipation.

In the limit of no dissipation, it is immediate to see that if the driving is 
shift-symmetric then the system is invariant under the transformation 
$\hat{S}_1$, and current generation is forbidden. If the the driving is
symmetric under time reversal, then the system is invariant under the 
transformation $\hat{S}_2$, and once again directed motion is forbidden.

We now carry further the symmetry analysis for a specific form of driving.
We consider the case of a bi-harmonic driving force:
\begin{equation}
F(t)=A\cos(\omega t)+B\cos(2\omega t+\phi)~.
\label{eq:biharmonic}
\end{equation}

For $A,B\neq 0$ the presence of both an even and an odd harmonic breaks the
shift symmetry $\hat{F}_{sh}$, independently of the relative value of the phase
$\phi$. On the other hand, whether the $\hat{F}_s$ symmetry is broken depends on
value of the phase $\phi$: for $\phi=n\pi$, with $n$ integer, the symmetry
$F_s$ is preserved, while for $\phi\neq n\pi$ it is broken. Therefore for
$\phi = n\pi$ current generation is forbidded, while for $\phi\neq n\pi$
it is allowed. Perturbative calculations \citep{flach00} show that the
average current of particles is, in leading order, proportional to $\sin\phi$,
in agreement with the above symmetry considerations.

We now consider the case of weak, nonzero dissipation. For
the sake
of simplicity, we restrict our analysis to the case of a bi-harmonic driving
of the form of Eq.~(\ref{eq:biharmonic}). As already mentioned the
shift-symmetry is broken as the driving consists both of even and odd
harmonics. Consider now the symmetry under time-reversal. For $\phi=n\pi$,
with $n$ integer, the driving has $\hat{F}_s$ symmetry. However, the system is not
symmetric under the transformation $\hat{S}_2$ because of dissipation.
Therefore the generation of a current is not prevented, despite the symmetry
of the driving. It was shown \citep{flach01} that the generated
current $I$ still shows an approximately sinusoidal dependence on the phase
$\phi$, but acquires a phase lag $\phi_0$: $I\sim \sin(\phi-\phi_0)$. Such a
phase lag corresponds to the dissipation-induced symmetry breaking.

\subsection{The quasiperiodically driven rocking ratchet}\label{sec:quasi}

We now consider the case of quasiperiodic driving. We consider a 
generic driving with two frequencies $\omega_1$, $\omega_2$. 
Quasiperiodic driving corresponds to an irrational value of the
ratio $\omega_2/\omega_1$. In order to analyze the relationship between
symmetry and transport in the case of a quasiperiodic driving, the
two phases
\begin{eqnarray}
\Psi_1 = \omega_1 t\\
\Psi_2= \omega_2 t
\end{eqnarray}
can be treated as {\it independent} variables \citep{pikovsky}. The symmetries
valid in the case of a perioding driving can then be generalized to the case
of a quasiperiodic ac force \citep{acta}.

The driving force $F(t)$ is said to be shift-symmetric, as for the periodic 
driving case of Sec. \ref{sec:periodic}, if it changes sign
under one of these transformations:
\begin{equation}
\Psi_{i} \to \Psi_{i}+\pi
\end{equation}
where $i$ is any subset of $\{1,2\}$, i.e. the $\pi$ shift is applied
to either any of the two variables, of to both of them. If $F$ is
shift-symmetric, then the system is invariant under the generalized symmetry
\begin{equation}
\tilde{S}_1: x\to -x,~~ \Psi_{i}\to \Psi_{i}+\pi
\end{equation}
and directed motion is forbidden.

The driving is said to be symmetric if
\begin{equation}
F(-\Psi_1+\chi_1,-\Psi_2+\chi_2) = F(\Psi_1,\Psi_2)
\end{equation}
with $\chi_1$, $\chi_2$ appropriately chosen constants. If the driving
is symmetric, in the dissipationless limit the system is invariant under the
generalized symmetry
\begin{equation}
\tilde{S}_2: x\to x,~~\Psi_j\to -\Psi_j+\lambda_j~~~(j=1,2)
\end{equation}
and directed transport is forbidden.

The two symmetries are the direct generalization of the symmetries for the
periodic case, and control directed motion in the case of a quasiperiodic
driving.

\subsection{The gating ratchet}\label{sec:gating}

In the {\it gating} ratchet \citep{nori04,borromeo05,borromeo}, particles 
experience an oscillating
potential which is spatially symmetric. A zero-average and time-symmetric
ac force is also applied. A current can be generated following a gating
effect, with the lowering of the potential barriers synchronized with
the motion produced by the additive force. This mechanism has to be
contrasted with the previously discussed ac-driven ratchets
with additive bi-harmonic driving, in which the
underlying mechanism is harmonic mixing \citep{marchesoni}.

The symmetry analysis for the gating ratchet was carried out in 
\cite{gommers08}, by following the generale procedure described 
in Sec.~\ref{sec:general}.  Consider a weakly damped
particle in an amplitude modulated symmetric potential $V(x)[1+m(t)]$. A
rocking force $F(t)$ is also applied. The Langevin equation for the
particle of mass $M$ is:
\begin{equation}
M\ddot{x}+\gamma\dot{x}=-V'(x)[1+m(t)] + F(t)+\xi(t)~.
\label{eq:langevin2}
\end{equation}
Both the amplitude modulation $m(t)$ and the rocking force $F(t)$ are
single-harmonic fields:
\begin{eqnarray}
m(t) &=& m_0\cos(\omega_1 t) \label{eq:mult}\\
F(t) &=& F_0\cos(\omega_2 t+\phi)~. \label{eq:add}
\end{eqnarray}
For the symmetry analysis, the noise term $\xi(t)$ can be ignored as it is
symmetric. Moreover, the dissipationless limit ($\gamma=0$) is considered first,
and a weak dissipation can be accounted for by an additional phase lag, as
discussed previously. 
The aim of the symmetry analysis is to
determine the conditions for the Langevin equation, Eq.~(\ref{eq:langevin2}),
to be invariant under the transformations $\hat{S}_1$, $\hat{S}_2$ (Eqs.~
\ref{eq:s1},\ref{eq:s2}).  The invariance of the Langevin equation under
$\hat{S}_1$ and/or $\hat{S}_2$ then prevents directed motion.

A relevant quantity for the symmetry analysis is the ratio between the driving
frequencies $\omega_1$, $\omega_2$. Limiting ourselves to the case of periodic
driving, we express the frequency ratio as $\omega_2/\omega_1=p/q$, with
$p, q$ co-primes. It is straightforward then to show that
the Langevin equation is invariant under the transformation $\hat{S}_1$ if
$q$ is even. Consider now the invariance under the transformation
$\hat{S}_2$. Elementary calculations show that the system is invariant
under $\hat{S}_2$ for $q\phi=n\pi$ with $n$ integer, and we therefore expect
a current $I$ of the form $I\sim\sin(q\phi)$. The symmetry
analysis of \cite{gommers08} shows that we should expect no current for $q$ even, and a current of
the form $I\sim\sin(q\phi)$ for $q$ odd. These results were obtained in the
dissipationless limit. It is straightforward now to take into account the
effects of weak dissipation.
Dissipation does not affect the reasoning for
the symmetry $\hat{S}_1$, i.e we still expect a zero current for $q$ even.
On the other hand dissipation breaks the invariance under the time-reversal
transformation $\hat{S}_2$, and a current can be generated also for
$q\phi=n\pi$.  We then expect a current of the form $I\sim\sin(q\phi+\phi_0)$,
with the effects of dissipation being accounted for by the phase lag $\phi_0$.

\section{Cold atom ratchets} \label{sec:coldatoms}

The first experiment on the ratchet effect using cold atoms in an optical 
lattice was reported by \cite{cecile}. In that work directed motion was 
observed in a spatially asymmetric {\it undriven} dark optical lattice. The 
ratchet effect with an undriven optical lattice was later on also demonstrated
for the case of spatially symmetric and shifted potentials 
\citep{kastberg1,kastberg2,kastberg3}.

The present work reviews the experimental work on the ratchet effect with 
{\it driven} optical lattices. In these experiments ac-drivings are applied,
either additively or multiplicatively, to drive the system out of equilibrium
and to break the relevant symmetries.

Before entering into the details of the realization of the ac-driven ratchets,
we summarize the basics of dissipative optical lattices and the underlying 
Sisyphus cooling mechanism. We refer to \citep{robi} for a more comprehensive
review of optical lattices.

\subsection{Dissipative optical lattices}\label{sec:lattice}

Optical lattices are periodic potentials for atoms created by the interference
of two or more laser fields. In near-resonant optical lattices a set of laser
fields produce at once the periodic potential acting on the atoms and the 
cooling mechanism, named Sisyphus cooling, which decreases their kinetic 
energy. The atoms are finally trapped at the bottom of the potential wells. 
We describe here the principles of these optical lattices in the case of a 
one-dimensional configuration and a $J_g=1/2\to J_e=3/2$ atomic transition. 
This is the simplest configuration in which Sisyphus cooling takes place.

%%%%%%%%%%%%%%%%%%%%%%%%%%%%%%%%%%%%%%%%%%%%%%%%%%%%%%%%%%%%%%%%%%%%%%%%%%%
\begin{figure}[ht]
\begin{center}
\mbox{\epsfxsize 3.in \epsfbox{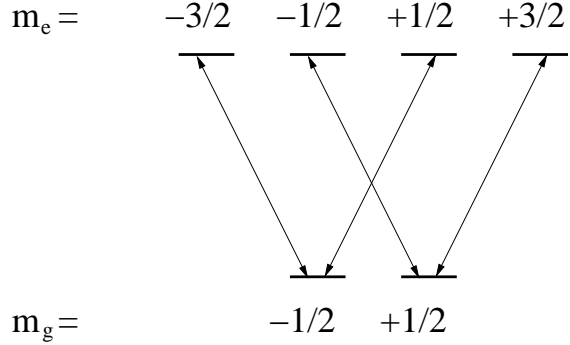}}
\end{center}
\caption{Atomic level scheme for a $J_g=1/2\to J_e=3/2$ transition. The arrows
indicate the couplings due to $\sigma^{+}$, $\sigma^{-}$ laser excitation.}
\label{fig5}
\end{figure}
%%%%%%%%%%%%%%%%%%%%%%%%%%%%%%%%%%%%%%%%%%%%%%%%%%%%%%%%%%%%%%%%%%%%%%%%%%%

\begin{figure}[ht]
\begin{center}
\mbox{\epsfxsize 3.in \epsfbox{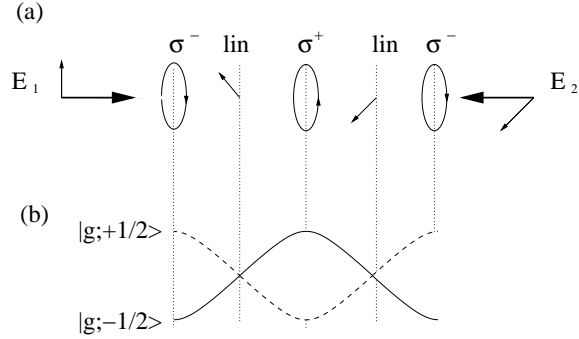}}
\end{center}
\caption{(a) Arrangement of laser fields in the so-called lin$\perp$lin
configuration, and resulting gradient of ellipticity. (b) Light shift of the
two ground-state Zeeman sublevels $|g,\pm 1/2\rangle$.}
\label{fig6}
\end{figure}

Consider a transition $J_g=1/2\to J_e=3/2$ (Fig.~\ref{fig5}) coupled to two
laser fields with the same amplitude and the same wavelength $\lambda$, linearly
polarized and counterpropagating. These laser fields are detuned below atomic
resonance and have orthogonal linear polarization (lin$\perp$lin configuration,
see Fig.~\ref{fig6}(a)):
\begin{equation}
\vec{E}_1(z,t) = \frac{1}{2}\vec{\epsilon}_x E_0 \exp [i(kz-\omega t)]+c.c
\end{equation}
\begin{equation}
\vec{E}_2(z,t) = \frac{1}{2}\vec{\epsilon}_y E_0 \exp [i(-kz-\omega t+\alpha)]+c.c
\end{equation}
where $\vec{\epsilon}_{x,y}$ are the unit vectors of linear polarization along 
the $(x,y)$ axes and $k=2\pi/\lambda$ and $\omega=kc$ are the laser field 
wavevector and angular frequency, respectively. The total electric field is
\begin{equation}
\vec{E}_1(z,t)+\vec{E}_2(z,t) = [E_{+}(z)\vec{\epsilon}_{+}+
E_{-}(z)\vec{\epsilon}_{-}]\exp (-i\omega t)+c.c.
\end{equation}
where $\vec{\epsilon}_{\pm}$ are the unit vectors of circular polarization. 
After elimination of the relative phase $\alpha$ through an appropriate
choice of the origin of the space- and time-coordinates, $E_{+}$ and
$E_{-}$ are given by:

\begin{eqnarray}
E_{+} &=& -i\frac{E_0}{\sqrt{2}} \sin kz ~,\\
E_{-} &=& \frac{E_0}{\sqrt{2}} \cos kz~.
\end{eqnarray}\label{eq_pol}
The superposition of the two laser fields $E_1$, $E_2$ produces therefore an
electric field characterized by a constant intensity and a spatial gradient of
polarization ellipticity of period $\lambda/2$, as shown in Fig.~\ref{fig6}(a).

%%%%%%%%%%%%%%%%%%%%%%%%%%%%%%%%%%%%%%%%%%%%%%%%%%%%%%%%%%%%%%%%%%%%%%%%%%%
%%%%%%%%%%%%%%%%%%%%%%%%%%%%%%%%%%%%%%%%%%%%%%%%%%%%%%%%%%%%%%%%%%%%%%%%%%%

We examine now the effects that the laser fields have on the atoms. The basic
mechanism responsible for the generation of a periodic potential is the
"light shift": a laser field coupling a given transition, and characterized
by an intensity $I_L$ and detuning $\Delta$ from atomic resonance, leads to
a shift of the ground state energy ("light shift") proportional to $I_L/\Delta$.

In the present case of a $J_g=1/2\to J_e=3/2$ transition there are two laser
fields coupling each ground state sublevel to the excited state, and 
contributions from all these couplings have to be taken into account to derive 
the light shifts $U_{\pm}$ for the ground state Zeeman sublevels
$|g,\pm 1/2\rangle$. We will omit here the details of the calculations, and 
simply report the final results for the light shifts (see \cite{robi} for the
derivation):
\begin{eqnarray}
U_{+} &=& 2\hbar\Delta_0' \left( \frac{I^{+}_L}{I_L} + \frac{I^{-}_L}{3I_L} \right)~,\\
U_{-} &=& 2\hbar\Delta_0' \left( \frac{I^{-}_L}{I_L} + \frac{I^{+}_L}{3I_L} \right)~.
\end{eqnarray}
Here $I^{\pm}_L=|E^{\pm}|^2$ are the intensities of the right- and
left-polarization components of the light, and $I_L=I^{-}_L+I^{+}_L$ is the total
intensity. The quantity $\Delta_0'$ is the {\it light shift per beam} for an optical 
transition with a Clebsch-Gordan coefficient equal to 1: 
\begin{equation}
\Delta_0'=\Delta\frac{\Omega_R^2/4}{\Delta^2+\Gamma^2/4}.
\end{equation}
Here $\Delta$ is the detuning of the optical field from atomic resonance and $\Gamma$ the
linewidth of the atomic transition. $\Omega_R$ is the  Rabi frequency \citep{cctbook} 
produced by an electric field of amplitude $E_0$ driving the optical
transition supposing its Clebsch-Gordan coefficient equal to 1. The square of the resonant
Rabi frequency is proportional to the light intensity, so in the limit of not too small
detuning $\Delta$, we find that the light shift per beam scales as $I/\Delta$, as already
mentioned. By substituting the expressions (\ref{eq_pol}) for $E^{+}$, $E^{-}$, the light
shifts $U_{\pm}$ can be rewritten as:
\begin{eqnarray}
U_{\pm} = \frac{U_0}{2} [ -2 \pm \cos kz ]
\end{eqnarray}
with
\begin{equation}
U_0 = -\frac{4}{3}\hbar \Delta_0'
\end{equation}

the depth of the potential wells. We therefore conclude that the light
ellipticity gradient produces a periodic modulation of the light shifts
of the ground state Zeeman sublevels (Fig.~\ref{fig6}(b)). These periodic 
modulation acts as an optical potential for the atoms, and indeed these 
periodically modulated light shifts are usually referred to as {\it optical 
potentials}. These optical potentials can be characterized by their depth
$U_0$ or by the related angular vibrational frequency at the bottom of the 
well $\omega_v$. For a $J_g=1/2\to J_e =3/2$ atom, the relationship
between these two quantities is given by
\begin{equation}
\hbar\omega_v = 2\sqrt{E_r U_0},
\end{equation}
where $E_r = \hbar^2k^2/2M$ is the recoil frequency for an atom of mass $M$.

We turn now to the analysis of the cooling mechanism, the so-called Sisyphus
cooling \citep{sisifo}, which decreases the kinetic energy of the atoms and 
allows their trapping at the bottom of the wells of the optical potential.

%%%%%%%%%%%%%%%%%%%%%%%%%%%%%%%%%%%%%%%%%%%%%%%%%%%%%%%%%%%%%%%%%%%%%%%%%%%
\begin{figure}[ht]
\begin{center}
\mbox{\epsfxsize 3.in \epsfbox{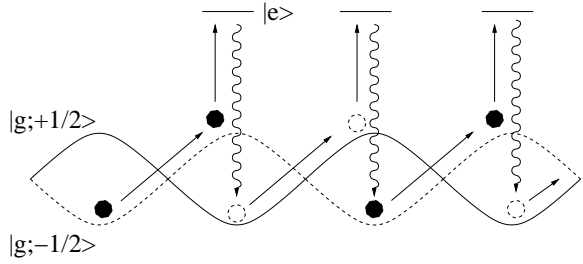}}
\end{center}
\caption{Sisyphus cooling mechanism.}
\label{fig7}
\end{figure}
%%%%%%%%%%%%%%%%%%%%%%%%%%%%%%%%%%%%%%%%%%%%%%%%%%%%%%%%%%%%%%%%%%%%%%%%%%%

Sisyphus cooling is determined by the combined action of the light shifts 
and of optical pumping, which transfers, through cycles of 
absorption/spontaneous emission, atoms from one ground state sublevel to the 
other one. This is illustrated in Fig.~\ref{fig7}. Consider an atom moving
with a positive velocity, and initially at $z=0$ in the state $|g,-1/2\rangle$.
While moving in the positive $z$ direction the atom climbs the potential curve
corresponding to its actual internal state. This has two consequences: first, a
part of the kinetic energy of the atoms is transformed in potential energy;
second, the component $\sigma^{+}$ of the light increases, which implies the
increase of the optical pumping rate towards the level $|g,+1/2\rangle$, i.e.
an increase of the probability of transfering the atom from the actual internal
state $|g,-1/2\rangle$ to the state $|g,+1/2\rangle$. At the top of the
potential hill ($z=\lambda/4$, see Fig.~\ref{fig6}) the polarization of the
light is purely $\sigma^{+}$, and the probability to transfer the atom into the
sublevel $|g,+1/2\rangle$ is very large. The transfer of the atom into the
level $|g,+1/2\rangle$ results into a loss of potential energy, which is
carried away by the spontaneously emitted photon. This process is repeated
several times, until the atom does not have enough energy any more to reach the
top of a potential hill, and it is trapped in a well. We notice here the
analogy with the myth of Sisyphus, king of Corinth, condemned forever to roll
a huge stone up a hill which repeatedly rolls back to the bottom before the
summit is reached. This is why the described cooling mechanism has been named
Sisyphus cooling. The described cooling process leads to the localization of 
the atoms at the bottom of the potential wells, and we obtain in this way an
{\it optical lattice}: an ensemble of atoms localized in a periodic potential.
We notice that the atoms are localized at the sites where their interaction 
with the light is maximum. It is because of this property that optical
lattices of this type are termed {\it bright} optical lattices.

An important quantity for the investigations reviewed in this work is the damping
rate of the atomic velocity ("cooling rate"). This will be the essential 
parameter to investigate the phenomenon of dissipation-induced symmetry breaking
in a rocking ratchet for cold atoms. Theoretical and experimental work
\citep{raithel,sancho} showed that the cooling rate is proportional to the scattering
rate $\Gamma^\prime$. For our 1D configuration and a $J_g=1/2\to J_e=3/2$ 
atom, the scattering rate can be expressed as:
\begin{equation}
\Gamma^\prime = \Gamma s_0 = \Gamma\frac{\Omega_R^2/4}{\Delta^2+\frac{\Gamma^2}{4}}~,
\end{equation}
where $s_0$ is the saturation per beam. Therefore the scattering rate $\Gamma^\prime$
will be used in the following to characterize the level of dissipation in the 
system under consideration.

\subsection{Rocking ratchet for cold atoms}

The realization of a driven ratchet requires essentially three elements. First,
a periodic potential; second, a fluctuating environment which results in
friction and in a fluctuating force. Finally, it should be possible to apply
a zero-mean ac-force to the particles (the atoms in the present case). All
these requirements can be satisfied by using cold atoms in optical lattices,
as it was demonstrated by \cite{schiavoni}. In that work the one-dimensional 
spatially symmetric lin$\perp$lin optical
lattice described in Sec.~\ref{sec:lattice} was taken as periodic potential. 

\begin{figure}[ht]
\begin{center}
\mbox{\epsfxsize 2.5in \epsfbox{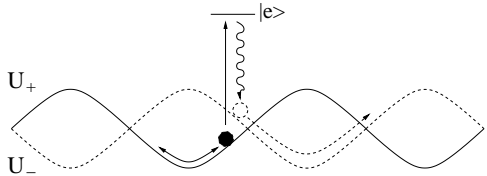}}
\end{center}
\caption{Stochastic process of optical pumping transferring, via an excited
state, an atom from a potential to the other one. The filled (empty) circle
represents the atom in the $|g,+1/2\rangle$ ($|g,-1/2\rangle$) ground state
sublevel.}
\label{fig8}
\end{figure}

We turn now to the analysis of the friction and fluctuations in the optical
lattice, the second element necessary to use optical lattices as a model
system for Brownian motors. As already discussed, the optical pumping between 
the different atomic ground state sublevels combined with the spatial
modulation of the optical potential leads to the cooling of the atoms and to
their localization at the minima of the optical potential. The essential fact
for the realization of Brownian motors is that even after the cooling phase,
characterized by a decrease of the kinetic energy of the atoms and their 
trapping in the optical potential, the atoms keep interacting with the light
fields and this induces fluctuations in the atomic dynamics. Indeed, consider 
an atom that has already lost enough energy to be trapped at the bottom of a
potential well.  The atom will then oscillate at the bottom of the well at 
angular frequency $\omega_v$. This situation is shown in Fig.~\ref{fig8}.

%%%%%%%%%%%%%%%%%%%%%%%%%%%%%%%%%%%%%%%%%%%%%%%%%%%%%%%%%%%%%%%%%%%%%%%%%%%
%%%%%%%%%%%%%%%%%%%%%%%%%%%%%%%%%%%%%%%%%%%%%%%%%%%%%%%%%%%%%%%%%%%%%%%%%%%

To be specific, consider
for example an atom initially in the $|g,+1/2\rangle$  state. Exactly at the
center of the well the light polarization is purely $\sigma^{+}$, which does
not allow the transfer from the $|g,+1/2\rangle$ state to the $|g,-1/2\rangle$
sublevel. However, out of the center of the well the light has also a nonzero
$\sigma^{-}$ component, which results in a nonzero probability to transfer
the atom from its original sublevel to the other one. Therefore the atom
can be transfered from one sublevel to the other one, and also the potential
experienced by the atom will change from $U_{+}$ to $U_{-}$, i.e. the force
experienced by the atom will change. As optical pumping is a stochastic process,
the (stochastic) transfer from a sublevel to the other one results in a
{\it fluctuating force}. Figure \ref{fig8} also shows how optical
pumping between different optical potentials leads to the transport of atoms
through the lattice: although the trapped atom does not have enough energy to
climb the potential hill, optical pumping allows the transfer from a potential
well to the neighbouring one. The optical pumping leads then to a {\it random walk}
of the atoms through the optical potential, and indeed normal diffusion has
been experimentally observed for an atomic cloud expanding in an optical
lattice \citep{regis}.

Two different quantities, the diffusion coefficients in momentum space $D_p$ 
and  in real space $D_{sp}$, can be introduced to characterize the atomic 
random walk 
in momentum and position respectively.

The momentum diffusion coefficient $D_p$, as determined by the fluctuations in
the dipole force, scales as $U_0^2/\Gamma^\prime$ \citep{sisifo}. The 
fluctuations in the dipole force are the main heating process in Sisyphus 
cooling. Thus the momentum diffusion coefficient determines, via the Einstein
relation $k_BT=D_p/\gamma$ with $\gamma$ the friction coefficient, the 
equilibrium temperature, which is found to be proportional to the potential
depth $U_0$ \citep{sisifo}.

The spatial diffusion coefficient $D_{sp}$ is instead predicted to be, for 
the range of lattice parameters corresponding to normal diffusion, approximately
proportional to the scattering rate $\Gamma^\prime$ \citep{robi}.

The last element necessary to implement a rocking ratchet is the oscillating
force. In order to generate a time-dependent homogeneous force, one of the
lattice beams is phase modulated, so to obtain the electric field
configuration:
\begin{equation}
\frac{1}{2} E_0 \left\{ \vec{\epsilon}_x \exp [i(kz-\omega t)] +
\vec{\epsilon}_y E_0 \exp [i(-kz-\omega t+\alpha (t) )]\right\} +c.c.~,
\end{equation}
where $\alpha (t)$ is the time-dependent phase. In the laboratory reference
frame this laser configuration generates a moving optical potential
$U[2kz-\alpha(t)]$. Consider now the dynamics in the moving reference frame
defined by $z'=z-\alpha(t)/2k$. In this accelerated reference frame the optical
potential is stationary. In addition to the potential the atom, of mass $m$,
experiences also an inertial force $F$ in the $z$ direction proportional to
the acceleration $a$ of the moving frame:
\begin{equation}
F=-Ma=\frac{M}{2k}\ddot{\alpha}(t)~.
\label{inertial}
\end{equation}
In this way in the accelerated frame of the optical potential the atoms
experience an homogeneous force which can be controlled by varying the phase
$\alpha (t)$ of one of the lattice beams. 

\subsection{Rocking ratchet with bi-harmonic driving}

The appropriate choice of the
phase $\alpha(t)$ for the realization of the spatially symmetric rocking
ratchet is
\begin{equation}
\alpha(t) = \alpha_0 \left[ \cos (\omega t) +\frac{\alpha_2}{4} \cos (2\omega t-\phi)
\right]
\label{alpha}
\end{equation}
with $\phi$ constant. Indeed, by using Eq.~(\ref{inertial}), we can see
immediately that in the accelerated frame of the optical potential the phase
modulation $\alpha(t)$ will result into a force
\begin{equation}
F=\frac{M\omega^2\alpha_0}{2k}
\left[ \cos (\omega t) + \alpha_2 \cos (2\omega t-\phi)
\right]
\label{force}
\end{equation}

with $\alpha_2$ the relative weight of the $2 \omega$ term.  This force 
is of the form needed for the realization of the spatially symmetric
rocking ratchet.

Experimentally, it is possible to obtain a phase modulation of the form
(\ref{alpha}) by simply using acousto-optical modulators and a set of radio-frequency
generators.  The exact technical realization is of no particular interest here,
and we refer to \cite{schiavoni} for further details. We only notice that
it is possible experimentally to carefully control the phase difference $\phi$
between the two harmonics. This allows us to carefully control the symmetry of
the system.

%%%%%%%%%%%%%%%%%%%%%%%%%%%%%%%%%%%%%%%%%%%%%%%%%%%%%%%%%%%%%%%%%%%%%%%%%%%
\begin{figure}[ht]
\begin{center}
\mbox{\epsfxsize 3.5in \epsfbox{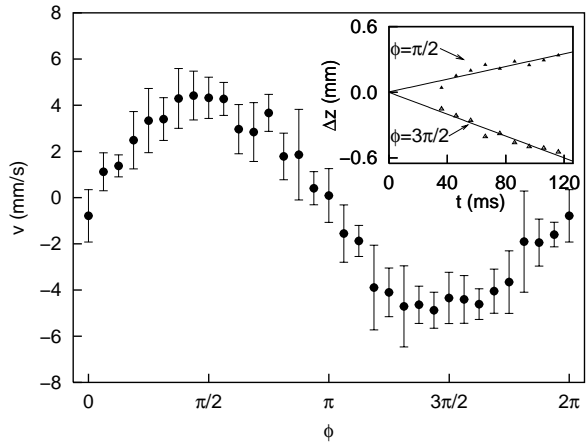}}
\end{center}
\caption{Average atomic velocity as a function of the phase $\phi$.
Inset: displacement of the center of mass of the atomic cloud as a function of
time for two different values of the phase $\phi$.  
[Reprinted figure with permission from \protect\citet{schiavoni}. 
Copyright 2003 of the American Physical Society.]}
\label{fig9}
\end{figure}
%%%%%%%%%%%%%%%%%%%%%%%%%%%%%%%%%%%%%%%%%%%%%%%%%%%%%%%%%%%%%%%%%%%%%%%%%%%

The experiment of \cite{schiavoni} on $^{85}$Rb atoms clearly demonstrated the 
control of the current through a spatially symmetric potential by varying the 
time-symmetries of the system. In that work, the dynamics of the atoms in the 
optical lattice was studied by direct imaging of the atomic cloud with a CCD 
camera. For a given phase $\phi$ the position of the center of mass of the 
atomic cloud was studied as a function of time. It should be noticed that in 
principle it is necessary to transform the measurements from the laboratory 
reference frame to the accelerated reference frame of the optical potential, by
using the coordinate transformation $z'=z-\alpha(t)/2k$. However in the case of 
\cite{schiavoni} this is not necessary as for the typical time scales of that 
experiment (period of the ac force and imaging time) the measured positions of 
the c.m. of the atomic cloud in the laboratory and in the accelerated reference
frame are approximately equal. The results of that experiment are reported in
Fig.~\ref{fig9}. It can be seen that the center of mass of the atomic cloud 
moves with constant velocity (see inset). This velocity shows the expected 
dependence on the phase $\phi$: for $\phi=n\pi$, with $n$ integer, the velocity
(current of atoms) is zero, while for $\phi=\pi/2, 3\pi/2$ the velocity reaches
a maximum (positive or negative). This because although the symmetry 
$F(t+T/2)=-F(t)$ is broken for any value of the phase $\phi$, there is a 
residual symmetry $F(t)=F(-t)$ which forbids the current generation. This 
symmetry is controlled by the phase $\phi$: for $\phi=n\pi$ it is realized, 
while for $\phi=(2n+1)\pi/2$ it is maximally broken.
\\
\\
The experiment of \cite{schiavoni} proved that the atoms can be set into 
directed motion through a symmetric potential by breaking the temporal symmetry
of the system.  That described experiment reproduced well the dependence of the
current on the phase $\phi$ derived in Sec.~\ref{sec:periodic} on the basis of
the analysis of symmetries which apply in the Hamiltonian limit, i.e. in the
absence of dissipation. This because \cite{schiavoni} performed the experiment
in the regime of relatively strong driving and small damping, which well 
approximates the Hamiltonian regime, as confirmed by detailed numerical 
simulations \citep{brown08}.

\subsubsection{Dissipation-induced symmetry breaking}

As discussed in Sec. \ref{sec:periodic}, the presence of weak damping results
in a shift of the curve representing the current as a function of the relative 
phase between the driving harmonics. This corresponds to a dissipation-induced
symmetry breaking, with the generation of a current for a system Hamiltonian 
symmetric in time and space. Such a ratchet regime was demonstrated experimentally
by \cite{gommers05b}.

In that experiment cesium atoms were loaded in a 3D optical lattice. 
A bichromatic driving force along one direction was applied by phase-modulating
one of the lattice beam. A rocking ratchet was realized in this way.
The level of dissipation was 
quantitatively characterized by the photon scattering rate $\Gamma'$, which 
can be controlled experimentally by varying the lattice fields parameters.

Different sets of measurements were performed for different values of the
scattering rate $\Gamma'$ at a constant depth of the optical potential.
This was done by varying simultanously the intensity $I_L$ and detuning
$\Delta$ of the lattice beams, so to keep the potential depth
$U_0\propto I_L/\Delta$ constant while varying the scattering rate
$\Gamma'\propto I_L/\Delta^2$. We notice that as $I_L$ and $\Delta$ can be
varied only within a finite range, dissipation cannot be suppressed completely,
i.e. it is not possible to obtain $\Gamma'=0$. However, as we will see, for the driving strength
considered in the experiment, the smallest accessible scattering rate results
in a phase shift which is zero within the experimental error, i.e. this choice
of parameters well approximates the dissipationless case. By then increasing
$\Gamma'$ it was possible to investigate the effects of dissipation.

%%%%%%%%%%%%%%%%%%%%%%%%%%%%%%%%%%%%%%%%%%%%%%%%%%%%%%%%%%%%%%%%%%%%%%%%%%%
\begin{figure}[ht]
\begin{center}
\mbox{\epsfxsize 3.in \epsfbox{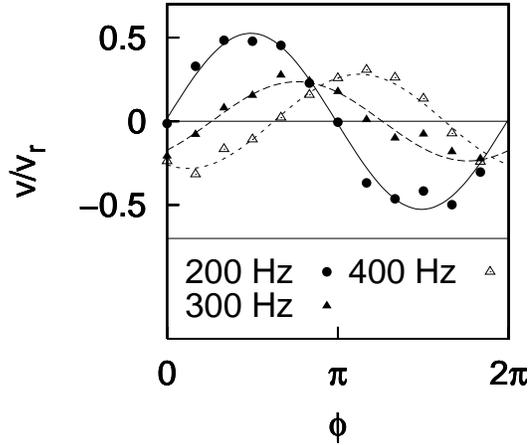}}
\end{center}
\caption{
Experimental results for the average atomic velocity, in units of the
recoil velocity $v_r=\hbar k/M$, as a function of the phase $\phi$. 
The recoil velocity is equal to 3.52 mm/s for the D$_2$ line of Cs atoms.
The lines are the best fit of the data with the function 
$v=v_{max}\sin(\phi-\phi_0)$. The optical potential is the same for 
all measurements, and corresponds to a vibrational frequency 
$\omega_v/(2 \pi)=170$ kHz. Different data sets correspond to different 
scattering rates obtained by varying the lattice detuning $\Delta$ and 
keeping constant the potential depth. The data are labeled by the quantity 
$\Gamma_s=[\omega_v/(2\pi)]^2/(\Delta/(2\pi))$ proportional to the scattering 
rate, reported in the bottom part. Driving parameters of the driving
are $\omega/(2 \pi)= 100$ kHz, $\alpha_0=27.2  {\rm rad}$, $\alpha_2=4$.
[Reprinted figure with permission from \protect\citet{gommers05b}. 
Copyright 2005 of the American Physical Society.]}
\label{fig_diss1}
\end{figure}

%%%%%%%%%%%%%%%%%%%%%%%%%%%%%%%%%%%%%%%%%%%%%%%%%%%%%%%%%%%%%%%%%%%%%%%%%%%
\begin{figure}[ht]
\begin{center}
\hspace*{-2.5cm}\mbox{\epsfxsize 4.in \epsfbox{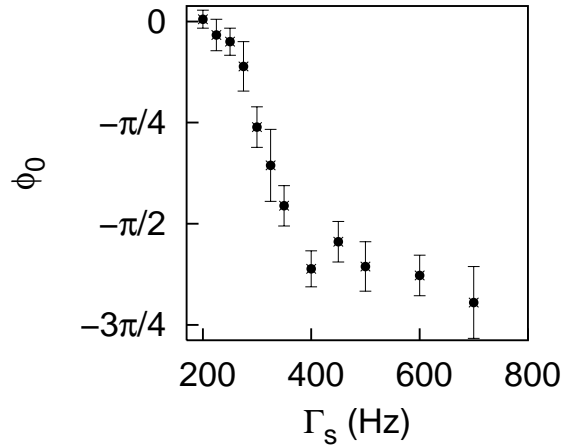}}
\end{center}
\caption{Experimental results for the phase shift $\phi_0$ as a function
of $\Gamma_s=[\omega_v/(2\pi)]^2/(\Delta/(2\pi)$, which is proportional to the
scattering rate. All the other parameters are kept constant, and are the
same as for Fig. \protect\ref{fig_diss2}. 
[Reprinted figure with permission from \protect\citet{gommers05b}. 
Copyright 2005 of the American Physical Society.]}
\label{fig_diss2}
\end{figure}

The results of the measurements of \cite{gommers05b}, reported in Fig.~\ref{fig_diss1}, demonstrate
clearly the phenomenon of dissipation-induced symmetry-breaking. In agreement
with previous theoretical work
\citep{flach00,flach01}, the measured current of atoms is well approximated
by $I_{max}\sin(\phi-\phi_0)$. Therefore, by fitting data as those reported in
Fig.~\ref{fig_diss1} with the function $v=v_{max} \sin(\phi-\phi_0)$
the phase shift $\phi_0$ was determined as a function of $\Gamma'$, as
reported in Fig.~\ref{fig_diss2}.  The measured phase shift $\phi_0$ is zero,
within the experimental error, for the smallest scattering rate examined in
the experiment.  In this case, no current is generated for $\phi=n\pi$, with
$n$ integer, as for this value of the phase the system is invariant under
time-reversal transformation.  The magnitude of the phase shift $\phi_0$
increases at increasing scattering rate, and differs significantly from zero.
The nonzero phase shift corresponds to current generation for $\phi=n\pi$,
i.e. when the system Hamiltonian is invariant under the time-reversal
transformation.  This result clearly demonstrates the breaking of the system
symmetry by dissipation.

%%%%%%%%%%%%%%%%%%%%%%%%%%%%%%%%%%%%%%%%%%%%%%%%%%%%%%%%%%%%%%%%%%%%%%%%%%%%%%
%%%%%%%%%%%%%%%%%%%%%%%%%%%%%%%%%%%%%%%%%%%%%%%%%%%%%%%%%%%%%%%%%%%%%%%%%%%%%%
\subsubsection{Rectification of fluctuations, current reversals and resonant 
activation in a system with broken Hamiltonian symmetry}
%%%%%%%%%%%%%%%%%%%%%%%%%%%%%%%%%%%%%%%%%%%%%%%%%%%%%%%%%%%%%%%%%%%%%%%%%%%%%%
%%%%%%%%%%%%%%%%%%%%%%%%%%%%%%%%%%%%%%%%%%%%%%%%%%%%%%%%%%%%%%%%%%%%%%%%%%%%%%

The cold atom experiments reviewed so far aimed to investigate the relationship
between symmetry and transport in rocking ratchets. In those experiments the 
generation of a current was studied as a function of the parameters controlling 
the symmetry of the system: the relative phase $\phi$, which controls the 
symmetry of the driving, and the scattering rate, which controls the 
symmetry-breaking of the system by dissipation.

However, in many other ratchet experiments, different aspects of
ratchets are investigated. Instead of studying the current as a function
of the symmetry-breaking parameters, a given investigation considers
thoroughly a  system with broken Hamiltonian symmetry. This can be realized, for 
example, by using a rocking ratchet with a spatially asymmetric potential 
or a temporally asymmetric force.  That study of the ratchet current amplitude 
as a function of the system parameters (driving amplitude and frequency, 
noise strength) reveals several distinguishing features of the ratchet 
effect. Namely, current reversals are  observed in correspondence of the 
variation of the driving amplitude and frequency. Furtermore, a non-monotonic
dependence of the amplitude of the generated current on the fluctuations 
level is a signature of the  rectification of fluctuations associated with 
the ratchet process.

%%%%%%%%%%%%%%%%%%%%%%%%%%%%%%%%%%%%%%%%%%%%%%%%%%%%%%%%%%%%%%%%%%%%%%%%%%%
\begin{figure}[ht]
\begin{center}
\mbox{\epsfxsize 3.5in \epsfbox{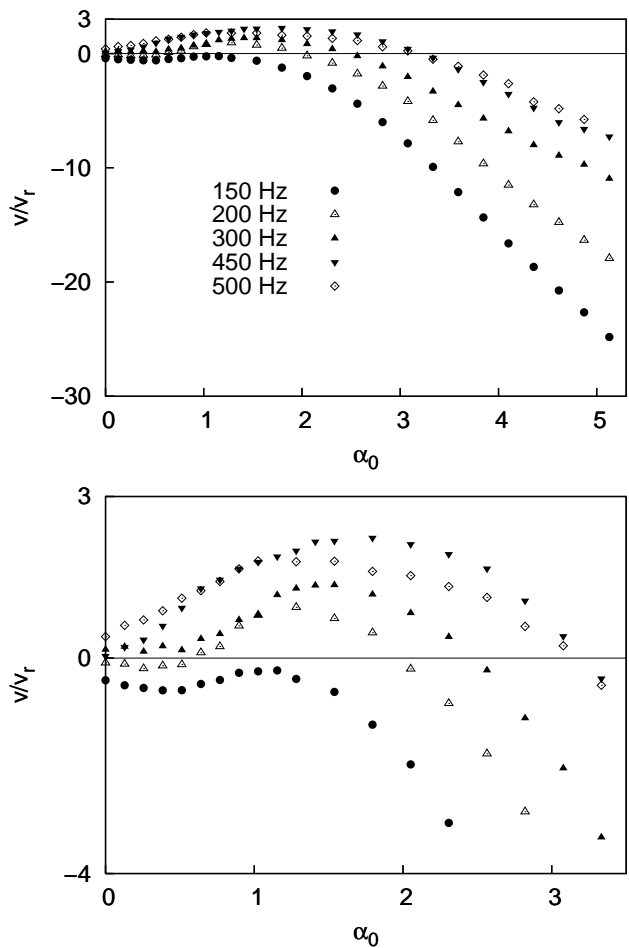}}
\end{center}
\caption{Experimental results for the atomic velocity as a function of the
amplitude of the phase modulation. The relative weight of the $2\omega$ term 
of the modulation (see Eq.~\ref{alpha}) is $\alpha_2=1$ for all data sets.
The top graph include all experimental
results, while the bottom graph evidences the region of small ac forces.
The optical potential is the same for all measurements. Different data
sets correspond to different optical pumping rate, and they are labeled
by $\Gamma_s=[\omega_v/(2\pi)^2]/(\Delta/(2\pi))$ ($\omega_v$ is the vibrational
frequency) which is proportional to the optical pumping rate. [Reprinted figure
with permission from \protect\citet{jones}. Copyright 2004 of the American Physical Society.]}
\label{fig:jones1}
\end{figure}
%%%%%%%%%%%%%%%%%%%%%%%%%%%%%%%%%%%%%%%%%%%%%%%%%%%%%%%%%%%%%%%%%%%%%%%%%%%

Investigations along these lines with cold atom ratchets with broken
Hamiltonian symmetry led to the observation of several hallmarks of the ratchet
effects. In the experiments by \cite{jones} and \cite{gommers05a}, a spatially
symmetric rocking ratchet with bi-harmonic driving was considered. Throughout
those investigations, the relative phase between the harmonics of the driving
was fixed to $\phi=\pi/2$, so that the Hamiltonian time-symmetry of the system
was broken.

In \cite{jones} the current of atoms through the lattice was studied as a 
function of the strength of the applied ac force for different values of the 
optical pumping rate $\Gamma^\prime$. Results of those measurements, reported in 
Fig.~\ref{fig:jones1}, show a clear dependence of the atomic current on the
amplitude of the applied force and on the optical pumping rate. Consider
first the dependence on the ac force magnitude. For a small amplitude of the
ac force the average atomic velocity is an increasing function of the force
amplitude, with the atoms moving in the positive direction. At larger amplitude
of the ac force the velocity decreases, and a current reversal is observed,
with the atomic cloud moving in the negative direction. This kind of behaviour,
named {\it current reversal}, is a hallmark of rocking ratchets. We examine 
now the dependence of the current on the optical pumping rate, i.e.  on the
noise level. We observe from Fig. \ref{fig:jones1} that such a dependence is 
very different depending on the ac force amplitude. For large amplitude of the
applied force the magnitude of the current (in absolute value) is a decreasing 
function of the optical pumping rate. This means that in this regime the motion 
can be attributed to deterministic forces and correspond to force rectification
by harmonic mixing: in a nonlinear medium the two harmonics, of frequency
$\omega$ and $2\omega$ and phase difference $\phi$, are mixed and the rectified
force produces a current  $I\sim\sin\phi$. In the considered experiment the 
nonlinearity of the medium is the anharmonicity of the optical potential. In 
this regime of rectification of the forces the noise does not play any 
constructive role in the generation of the current of atoms. On the contrary,
the noise disturbs the process of rectification of the forces, and indeed 
the current decreases for increasing optical pumping rate. Thus,
this regime does not correspond to the rectification of fluctuations. A very 
different dependence of the current amplitude on the optical pumping
rate was found in the regime of small amplitudes of the applied force. Indeed
in this regime the current is for small pumping rates an increasing function
of the pumping rate, and the current vanishes in the limit of vanishing optical
pumping rates. At larger pumping rates the current reaches a maximum and then
decreases again. This bell-shaped dependence of the current on the optical
pumping rate is a typical signature of a Brownian motor: in the absence of
fluctuations the current is zero, then increases until the fluctuations are
so large that the presence of the potential and of the applied fields become
irrelevant, and the current decreases again. Thus, in the regime of small ac 
force amplitude the optical lattice provides an implementation of a Brownian 
motor.

Another important parameter for the rectification mechanism is the 
driving frequency. Consider first the problem of the escape of a Brownian
particle from a {\it single} potential well. It is well known that in the
presence of nonadiabatic driving the lifetime of the particle in the well 
can be significantly reduced, a phenomenon named {\it resonant activation}
\citep{devoret1,devoret2,chaos}.  
Resonant activation has also been theoretically studied for Brownian particles 
in periodic potentials. Also in this case the nonadiabatic driving may result
in a significant enhancement of the activation rate. Moreover, whenever the 
spatio-temporal symmetry of the system is broken, the resonant activation gives 
then rise to {\it resonant} rectification of fluctuations 
\citep{dykman,goychuk,luchinsky2}. The resonant activation results in a 
resonance as function of the driving frequency in the current of atoms 
through the periodic potential. The theoretical work 
\citep{dykman,soskin,luchinsky2} also predicted that by changing the frequency 
of the driving it is possible to control the direction of the diffusion.

\begin{figure}[ht]
\begin{center}
\mbox{\epsfxsize 3.in \epsfbox{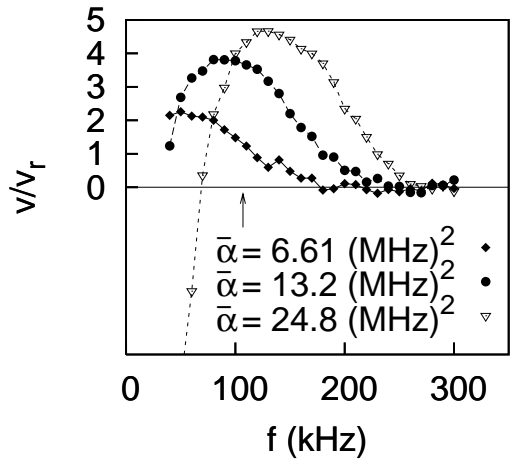}}
\end{center}
\caption{Experimental results for the average atomic velocity as a function
of the driving frequency $f=\omega/(2 \pi)$, for different amplitudes of the 
driving force.  As from Eqs.~(\ref{alpha},\ref{force}), for each data set the 
force is kept constant, while scanning the driving frequency, by varying the 
amplitude $\alpha_0$ of the phase modulation according to
$\alpha_0=\bar{\alpha}/f^2$.
The optical potential constant for all measurements corresponds
to a vibrational frequency $\omega_v/(2 \pi) =170$ kHz.
The driving frequency satisfying the condition $2\omega=\omega_v$ is
indicated by an arrow.  The values for the
velocity are expressed in terms of the recoil velocity $v_r$, equal to
$3.52$ mm/s for the Cs D$_2$ line. The relative weight of the $2\omega$
term of the modulation (see Eq.~\ref{alpha}) is $\alpha_2=1$ for all data sets.
The lines are guides for the eye. [Reprinted figure with permission from \protect\citet{gommers05a}. Copyright 2005 of the American Physical Society.]}
\label{fig_act}
\end{figure}

The experimental work by \cite{gommers05a} precisely studied the driving 
frequency dependence of the rectification mechanism in a periodically driven 
rocking ratchet for cold atoms. In that work, the current of atoms through the ac driven lattice 
was studied as a function of the driving frequency $\omega$, for a given relative phase 
$\phi=\pi/2$ between the driving harmonics, so to break the time-symmetry of
the system.  The build-up of a resonance was observed when the amplitude of 
the driving was progressively increased, as shown in Fig~\ref{fig_act}. The
resonance appears in the regime of non-adiabatic driving
($2\omega \gtrsim \omega_v$), and
a current reversal is observed on the low-frequency side of the resonance, in
agreement with the general theory \citep{dykman,soskin,luchinsky2}.

%%%%%%%%%%%%%%%%%%%%%%%%%%%%%%%%%%%%%%%%%%%%%%%%%%%%%%%%%%%%%%%%%%%%%%%%%%%

%%%%%%%%%%%%%%%%%%%%%%%%%%%%%%%%%%%%%%%%%%%%%%%%%%%%%%%%%%%%%%%%%%%%%%%%%%%

\subsection{Multi-frequency driving, and route to quasiperiodicity}

Experiments by
\cite{gommers06,gommers07} investigated the transition from
periodic to quasiperiodic driving, and examined how the symmetry
analysis is modified in this transition. In these experiments, a 
multifrequency driving was used, as 
obtained by combining signals at three different frequencies:
$\omega_1$, $2\omega_1$ and $\omega_2$. For $\omega_2/\omega_1$ irrational
the driving is quasiperiodic. Clearly, in a real experiment
$\omega_2/\omega_1$ is always a rational number, which can be written
as  $\omega_2/\omega_1=p/q$, with $p,q$ two coprime positive integers.
However, as the duration of the experiment is finite, by choosing
$p$ and $q$ sufficiently large it is possible to obtain a driving which
is effectively quasiperiodic on the time scale of the experiment.
Different forms of multifrequency driving were examined in the experimental
realizations, each probing a different symmetry.

The first form of driving examined by \cite{gommers06,gommers07} consisted of 
the sum of three harmonics:

\begin{equation}
F(t)=A\cos(\omega_1t)+B\cos(2\omega_1 t+\phi)+C\cos(\omega_2 t+\delta)~.
\label{eq:triharmonic}
\end{equation}

In the analysis, the effects of dissipation can be neglected, as we know that
it results in an additional phase shift. Consider first the case of periodic
driving, with $\omega_2/\omega_1$ rational. For
biharmonic driving, i.e. $C=0$ in Eq. (\ref{eq:triharmonic}), the shift symmetry is
broken for any value of $\phi$, while the time-reversal symmetry is preserved
for $\phi=n\pi$, with $n$ integer. A current of the form $I\sim \sin\phi$
is obtained as a result. Consider now the effect of the third harmonic, i.e.
$C\neq 0$ in Eq. (\ref{eq:triharmonic}. For a  phase $\delta=0$ of the 
$\omega_2$ harmonic, this additional
driving is invariant under time reversal, and therefore the total driving
is still invariant under time-reversal for $\phi=n\pi$. Instead, for
$\delta \neq 0$ the symmetry under time-reversal is broken and directed
transport is allowed also for $\phi=n\pi$. In other words, for $\delta\neq 0$
the third driving leads to an additional phase shift of the current as a
function of $\phi$. The magnitude of such a shift depends on the phase
$\delta$. Taking dissipation also into account, it follows that the
current will show the dependence $I\sim \sin(\phi-\phi_0)$ where $\phi_0$
includes the phase shift produced by dissipation and the phase shift
produced by the harmonic at frequency $\omega_2$.

We now turn to the case of a quasiperiodic driving, as obtained in the
case of irrational $\omega_2/\omega_1$. As discussed in Sec.
\ref{sec:quasi} the symmetry analysis for the periodic driving can be
generalized to the quasiperiodic case by treating the phases
$\Psi_1=\omega_1 t$ and $\Psi_2=\omega_2 t$ as independent variables.
We notice that the driving considered here, Eq. (\ref{eq:triharmonic}),
is invariant under the transformation $\Psi_2\to -\Psi_2+\chi_2$
for any $\delta$, as $\delta$ can be reabsorbed in $\chi_2$. Therefore
the invariance under the transformation $\tilde{S}_2$ is entirely
determined by the invariance of $F$ under the transformation
$\Psi_1\to -\Psi_1+\chi_1$; i.e., we recover the results for biharmonic
driving: $\tilde{S}_2$ is a symmetry, and therefore directed motion is
forbidden, for $\phi=n\pi$. Hence, in the quasiperiodic limit, the third
harmonic at frequency $\omega_2$ is not relevant for the symmetry of the
system, which is entirely determined by the biharmonic term at frequency
$\omega_1$, $2\omega_1$.

\begin{figure}[t]
\centering
\includegraphics*[width=.6\textwidth]{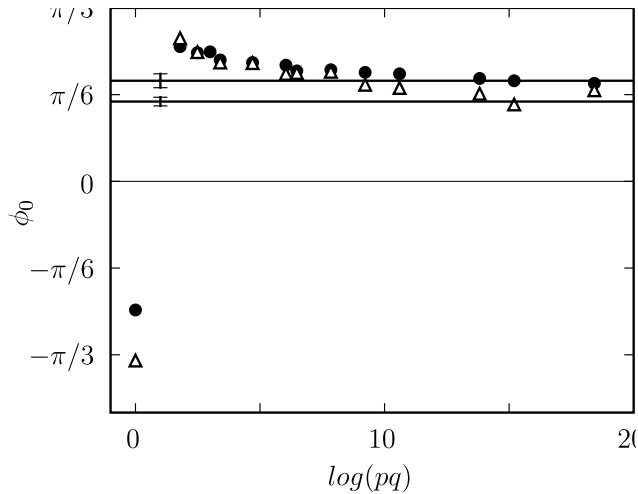}
\caption[]{Experimental results for the phase shift $\phi_0$ as a
function of $pq$ which characterize the degree of periodicity of the
driving. The two data sets, represented by open triangles and closed 
circles, correspond to different amplitudes of the driving. The two 
horizontal lines indicate the phase shift $\phi_0$ for biharmonic 
drive, i.e., in the absence of the driving at frequency $\omega_2$. 
[Reprinted figure with permission from \protect\citet{gommers06}. 
Copyright 2006 of the American Physical Society.]}
\label{fig:quasip1}
\end{figure}

In the experiment by \cite{gommers06}, the transition to quasiperiodicity 
was investigated by studying the atomic current as a function of $\phi$ 
for $\omega_2/\omega_1=p/q$ with $p$ and $q$ coprimes. By increasing $p$ and
$q$ the driving can be made more and more quasiperiodic on the finite
duration of the experiment, with the quantity $pq$ a possible measure of
the degree of quasiperiodicity. To verify the predictions of the symmetry
analysis, the average atomic current was measured as a function of $\phi$,
for different choices of $p$ and $q$. The data were fitted with the
function $v= v_{max}\sin(\phi-\phi_0)$. The resulting value for the phase
shift $\phi_0$ is plotted in Fig. \ref{fig:quasip1} as a function of
$pq$.

For small values of the product $pq$, i.e., for periodic driving, the
harmonic at frequency $\omega_2$ leads to a shift which strongly depends
on the actual value of $pq$. For larger values of $pq$, i.e., approaching
quasiperiodicity, the phase shift $\phi_0$ tends to a constant value. Such
a value was found to be independent of $\delta$, and coincides with the
phase shift $\phi_0$ measured in the case of pure biharmonic driving
(horizontal lines in Fig. \ref{fig:quasip1}), which is determined by
the finite damping of the atomic motion. The experimental results of
Fig. \ref{fig:quasip1} prove that, in agreement with the symmetry analysis, 
in the quasiperiodic limit the only
relevant symmetries are those determined by the periodic biharmonic
driving and by dissipation. For a driving of a form Eq. \ref{eq:triharmonic},
quasiperiodicity therefore restores the symmetries which hold in the absence
of the additional driving which produced quasiperiodicity.

A different form of multi-frequency driving was also examined by 
\cite{gommers06}. The driving force was obtained by multiplying the bi-harmonic
driving at frequencies $\omega_1$, $2\omega_1$ with the driving at frequency
$\omega_2$. This was done by applying to one of the lattice beams a frequency 
modulation of the form
\begin{equation}
\dot{\alpha}(t) = \alpha_0 \sin (\omega_2 t+\delta )\left[  \sin(\omega_1 t)+
\frac{\alpha_2}{4}\sin (2\omega_1 t)\right]
\end{equation}
which results into a force
\begin{eqnarray}
F(t)&=&-\frac{M\alpha_0}{k}\left\{ \omega_2 \cos(\omega_2 t+\delta)
\left[ \sin(\omega_1 t) + \frac{\alpha_2}{4} \sin(2\omega_1 t)\right] \right. \nonumber \\
&+&  \left. \omega_1 \sin(\omega_2 t+\delta)
\left[\cos(\omega_1 t) + \frac{\alpha_2}{2}\cos(2\omega_1 t\right] \right\}
\end{eqnarray}
It was shown that in this case quasiperiodicity results in the total suppression of
transport.

Consider first the case of periodic driving. We indicate, as before,
$\omega_2=(p/q)\omega_1$. The period $T$ of $F(t)$ is then $T=q T_1=pT_2$,
with $T_i=2\pi/\omega_i$ ($i=1,2$). Under the transformation $t\to t+T/2$ we
have: $\omega_1 t \to \omega_1 t+q \pi$,  $\omega_2 t \to \omega_2 t+p \pi$.
By replacing these transformations in $F(t)$ it is straighforward to see that
$F(t)$ satisfies the shift symmetry $F(t)=-F(t+T/2)$ if $q$ is even, and $p$ is
odd. In this case directed transport is forbidden. If instead this condition is
not satisfied, i.e. if $q$ is odd, directed transport is not forbidden. In this 
case directed transport is controlled by the $\hat{S}_2$ symmetry which is realized, 
in the dissipationless limit, if the driving $F(t)$ is symmetric under time-reversal.
The symmetry under time-reversal depends entirely on the phase $\delta$ of the 
driving at frequency $\omega_2$: for $q\delta=(n+1/2)\pi$, with $n$ integer, the 
driving is symmetric. Otherwise, the symmetry under time-reversal is broken. The 
current is expected to show a sinusoidal dependence on $q\delta-\pi/2$, and 
dissipation will account for an additional shift.

\begin{figure}[hbt]
\begin{center}
\mbox{\epsfxsize 3.in \epsfbox{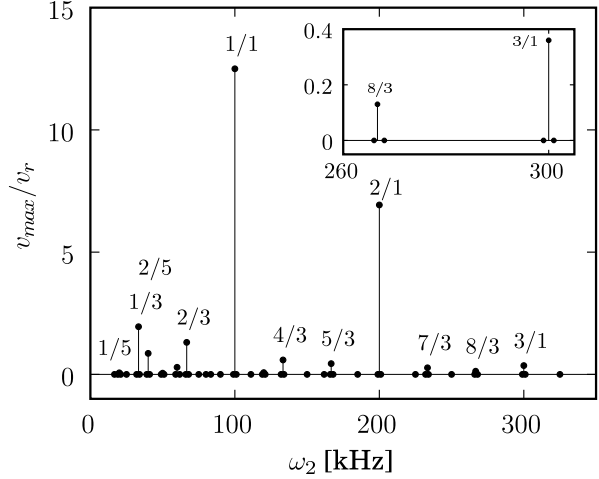}}
\end{center}
\caption{
Maximum average velocity as a function of the driving frequency $\omega_2$.
The data corresponding to a nonzero velocity are labelled by
$p/q=\omega_2/\omega_1$. The inset magnifies a portion of the plot.[Reprinted figure with permission from \protect\citet{gommers06}. Copyright 2006 of the American Physical Society.]}
\label{fig:quasip2}
\end{figure}

In the experiment, the average atomic velocity was measured as a function of
$\delta$ for different values of the driving frequency 
$\omega_2=(p/q)\omega_1$,with $p,q$ co-primes. By fitting the data with
$v=v_{max}\sin(q\delta-\delta_0)$,
the maximum velocity $v_{max}$ was determined as a function of $\omega_2$. 
The results of \cite{gommers06} , shown in Fig. \ref{fig:quasip2}, demonstrate 
the relationship between symmetry and transport, valid in the periodic case, 
discussed above. In fact, a current
was observed only for values of the ratio of driving frequencies
$\omega_2/\omega_1=p/q$ with $q$ odd, which is precisely the  requirement for
the shift symmetry to be broken.

%%%%%%%%%%%%%%%%%%%%%%%%%%%%%%%%%%%%%%%%%%%%%%%%%%%%%%%%%%%%%%%%%%%%%%%%%%%
\vspace{0.5cm}
%%%%%%%%%%%%%%%%%%%%%%%%%%%%%%%%%%%%%%%%%%%%%%%%%%%%%%%%%%%%%%%%%%%%%%%%%%

%%%%%%%%%%%%%%%%%%%%%%%%%%%%%%%%%%%%%%%%%%%%%%%%%%%%%%%%%%%%%%%%%%%%%%%%%%%
\vspace{0.75cm}
\begin{figure}[hbt]
\begin{center}
\mbox{\epsfxsize 3.in \epsfbox{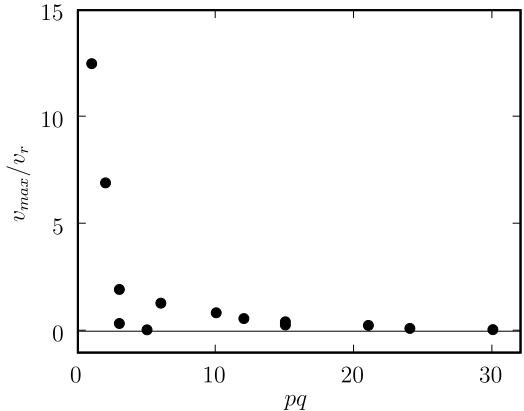}}
\end{center}
\caption{
Maximum average velocity as a function of $pq$, where $p$ and $q$ are
the co-primes defined by the ratio of the driving frequencies:
$p/q=\omega_2/\omega_1$. [Reprinted figure with permission from \protect\citet{gommers06}. Copyright 2006 of the American Physical Society.]}
\label{fig:quasip3}
\end{figure}
%%%%%%%%%%%%%%%%%%%%%%%%%%%%%%%%%%%%%%%%%%%%%%%%%%%%%%%%%%%%%%%%%%%%%%%%%%

Consider now the case of quasiperiodic driving. To analyze
this case, we introduce the two variables $\psi_1=\omega_1 t$ and
$\psi_2=\omega_2 t$, to be treated as independent, and consider the generalized
symmetries $\tilde{S}_1$, $\tilde{S}_2$. It is immediate to verify that
$F$ changes sign under the transformation $\psi_2\to \psi_2+\pi$, i.e.
$F$ is shift symmetric with respect to $\psi_2$. It follows that the
system is invariant under the generalized symmetry $\tilde{S}_1$.
Directed transport is therefore forbidden. In order to study the 
transition to quasiperiodicity,  the data of Fig.~\ref{fig:quasip2} were 
re-arranged as a function of $pq$ which characterizes the quasiperiodic 
character of the driving on the finite duration of the experiment.
The results are shown in Fig.~\ref{fig:quasip3}. It appears that for 
large $pq$ values the amplitude of the atomic current decreases to zero. 
This demonstrates that directed transport is destroyed in the quasiperiodic 
limit, as a results of  the restoration of the shift symmetry of the driving.

\subsection{Gating ratchet}

As discussed in Sec.~\ref{sec:gating}, in a gating ratchet particles 
experience an amplitude-modulated potential which is spatially symmetric. 
A zero-average and time-symmetric ac force is also applied. A current
can be generated following a gating effect, with the lowering of the 
potential barriers synchronized with the motion produced by the 
additive force. 

A gating ratchet for cold atoms was demonstrated experimentally by 
\cite{gommers08}. The ratchet was realized with cold rubidium atoms
in a driven 1D dissipative optical lattice. A single-harmonic periodic
modulation of the potential depth was applied, together with a single
harmonic rocking force. As in Sec.~\ref{sec:gating}, the frequencies of the
multiplicative (potential modulation) and additive (rocking force) drivings 
are denoted with $\omega_1$ and $\omega_2$, respectively, with the relative 
phase indicated by $\phi$.

\begin{figure}[t]
\begin{center}
\includegraphics[width=3.in]{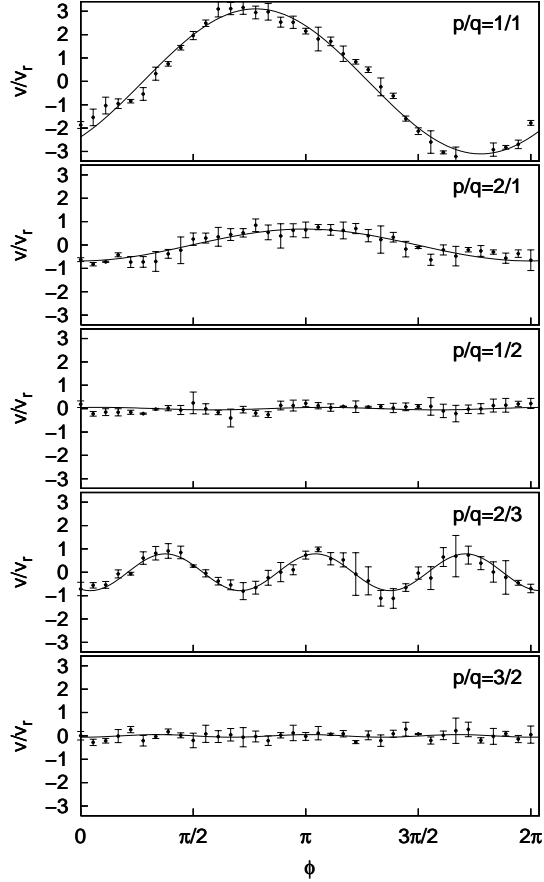}
\end{center}
\caption{Experimental results for a gating ratchet for cold atoms. The
average atomic velocity is reported as a function of the phase offset
$\phi$ between multiplicative and additive drivings. The atomic velocity is
expressed in terms of the recoil velocity $v_r$, which for $^{87}$Rb is
equal to $5.9$ mm/s.
Different data sets correspond to different values of the frequency $\omega_2$
of the additive (rocking) force. The frequency of the multiplicative driving
is the same for all data sets, and it is equal to $150$ kHz. The data sets
are labelled by the ratio $p/q=\omega_2/\omega_1$. 
% For all the measurements, the intensity modulation index is $91$ \%.
% The intensity per lattice beam is $I_L=22$ mW/cm$^2$.
% The detuning of the lattice fields from resonance is
% $\Delta =-15 \Gamma$. The frequency modulation depth $f_0$
% is $f_0= 750$ kHz for all the data sets.
The lines are the best fits of the data with the function
$v=v_{\rm max}\sin(q\phi+\phi_0)$. [Reprinted figure with permission 
from \protect\citet{gommers08}. Copyright 2008 of the American Physical Society.]}
\label{fig:gating1}
\end{figure}

\begin{figure}[t]
\begin{center}
\includegraphics[width=3.in]{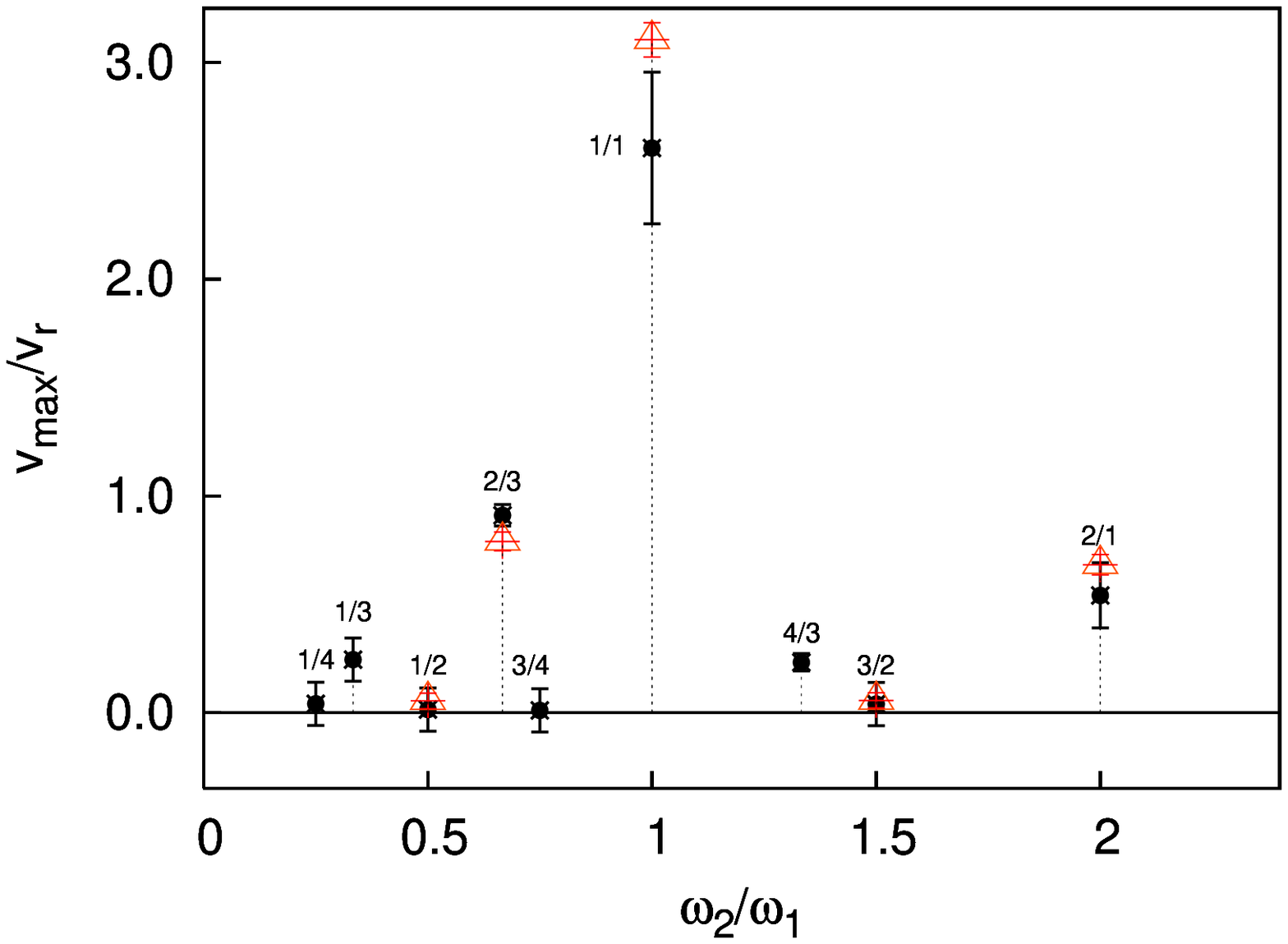}
\end{center}
\caption{Experimental results for the atomic current amplitude
as a function of the frequency ratio $\omega_2/\omega_1$, as obtained by 
fitting data as those in Fig.~\ref{fig:gating1} with the function
$v=v_{\rm max} \sin(q\phi+\phi_0)$. The triangles represent the fit of the
data of Fig.~\ref{fig:gating1}, the circles the fit of the data taken during
a different measurement session. 
[Reprinted figure with permission from \protect\citet{gommers08}.
Copyright 2008 of the American Physical Society.]}
\label{fig:gating2}
\end{figure}

The results of \cite{gommers08} are reported in Fig.~\ref{fig:gating1} and
\ref{fig:gating2}. In Fig.~\ref{fig:gating1} the average atomic velocity is
reported as a function of the phase offset $\phi$. Different data sets
were taken for different values of the ratio $\omega_2/\omega_1$.
Figure \ref{fig:gating2} reports the corresponding current amplitude.

The experimental results of Figs.~\ref{fig:gating1} and \ref{fig:gating2}  
constitute the experimental demonstration of a gating ratchet for
cold atoms. The presence of both a single-harmonic additive driving and a
single-harmonic multiplicative driving allows the breaking of the symmetries
of the system, and a current is generated as a result. 

The observations of \cite{gommers08} are in agreement with the symmetry 
analysis of Sec.~\ref{sec:gating}. In fact, the analysis of the data for 
the different values of the driving frequencies ratio 
$\omega_2/\omega_1=p/q$ shows that a current is generated only for $q$ odd,
as also evidenced in Fig.~\ref{fig:gating2}, and in this case the average
atomic velocity exhibits a dependence on the phase $\phi$ of the form 
$v=v_{\rm max}\sin(q\phi+\phi_0)$.

As already pointed out in Sec.~\ref{sec:gating}, there is an important 
difference  between the gating ratchet realized by \cite{gommers08}
and the previously demonstrated rocking ratchet with additive bi-harmonic
driving \citep{schiavoni}. In the rocking ratchet the underlying
mechanism is harmonic mixing \citep{marchesoni}, while the gating ratchet
relies on a gating effect, with the lowering of the potential barriers
synchronized with the motion produced by the additive force. This important
difference is also manifest in the different conditions for the generation
of a current. For example, in the gating ratchet a large current can be
obtained when the two driving frequencies are equal, while the rocking ratchet
requires harmonic mixing of two different frequencies.

\section{Outlook} \label{sec:outlook}

This article reviewed recent experimental realization of ac driven
ratchets with cold atoms in driven optical lattices. 

Such a system allowed to realize experimentally rocking and gating ratchets,
and to precisely investigate the relationship between symmetry and transport 
in these ratchets, both for the case of periodic and quasiperiodic driving.

The extreme tunability of optical lattices offers an unique possibility to
investigate further the ratchet effect. For example, 2D and 3D optical lattices
can be used to investigate complex multi-dimensional rectification mechanism
\citep{denisov}.

Disordered potentials and/or time-forces may be exploited to study the role of
disorder in the transport in a ratchet device \citep{harms,marchesoni97}. 
Cold atoms in optical lattices may also allow for the realization of a 
quantum ratchet \citep{reimann97}, where the transport is produced
by the interplay between tunneling and dissipation. Finally, the use of a
Bose-Einstein condensate could allow to model the ratchet effect for vortices.
By using multi-dimensional ratchet set-ups, as those proposed by \cite{denisov},
it should be possible to create vorticity in a controlled way. The very same
ratchet set-up could then allow to control the vortex motion.
This would constitute a clean model system for superconductor physics. 

% The Appendices part is started with the command \appendix;
% appendix sections are then done as normal sections
% \appendix

% \section{}
% \label{}

% Bibliographic references with the natbib package:
% Parenthetical: \citep{Bai92} produces (Bailyn 1992).
% Textual: \citet{Bai95} produces Bailyn et al. (1995).
% An affix and part of a reference:
%   \citep[e.g.][Ch. 2]{Bar76}
%   produces (e.g. Barnes et al. 1976, Ch. 2).

% \bibitem[Names(Year)]{label} or \bibitem[Names(Year)Long names]{label}.
% (\harvarditem{Name}{Year}{label} is also supported.)
% Text of bibliographic item

\end{document}